\documentclass[12pt]{article}
\usepackage{amsmath, amsfonts,euscript,bbm}
\usepackage{epsfig, cite}
\usepackage{color}
\usepackage{ifthen}
\usepackage{graphicx}
\usepackage{setspace} 
\newboolean{Notes}\setboolean{Notes}{true}
\newcommand{\notes}[1]{\ifthenelse{\boolean{Notes}}{\textcolor{black}{#1}}{}}

\newcommand{\skyp}[1]{}

\begin{document}

\bigskip
\hskip 4in
\vbox{\baselineskip12pt \hbox{FERMILAB-PUB-05-153-A}}
\bigskip\bigskip

\centerline{\Large Gravity from a Modified Commutator}
\bigskip
\bigskip
\bigskip
\centerline{\bf Mark G. Jackson}
\medskip
\centerline{Particle Astrophysics Center}
\centerline{Fermi National Accelerator Laboratory}
\centerline{Batavia, Illinois 60510}
\centerline{\it markj@fnal.gov}
\bigskip
\bigskip
\begin{abstract}
We show that a suitably chosen position-momentum commutator can elegantly describe many features of gravity, including the IR/UV correspondence and dimensional reduction (`holography').  Using the most simplistic example based on dimensional analysis of black holes, we construct a commutator which qualitatively exhibits these novel properties of gravity.  Dimensional reduction occurs because the quanta size grow quickly with momenta, and thus cannot be ``packed together" as densely as naively expected.  We conjecture that a more precise form of this commutator should be able to quantitatively reproduce all of these features.
\end{abstract}
\bigskip
\bigskip
\bigskip
\bigskip
\bigskip
\bigskip
\bigskip
\bigskip
\centerline{{\em This essay received an Honorable Mention in the}}
\centerline{{\em 2005 Gravity Research Foundation Essay Competition}}
\newpage            
\baselineskip=18pt
\doublespacing
\section{Introduction}
The most intriguing aspect of gravity is that of holography.  It was 't Hooft \cite{'tHooft:1993gx} who first observed that the true degrees of freedom in space are on the boundary and not in the volume as naively expected.  More precisely, it was shown that a system with gravity should be describable by a quantum theory in one less dimension.  Subsequent developments by Susskind \cite{Susskind:1994vu} worked out more consequences of this dimensional reduction, or the more fashionable name `holography', and Maldacena \cite{Maldacena:1997re} conjectured this may be demonstrated in string theory based on an explicit construction in anti-deSitter space.

This is not the first time that a vast reduction from the expected degrees of freedom has been discovered.  In their seminal article on string thermodynamics, Atick and Witten~\cite{Atick:1988si} found that a 10- or 26-dimensional string theory at high temperature behaved more like a 2-dimensional field theory.  In their speculation on why this might occur they comment:

\singlespacing
\begin{quotation}
\small{``...Let us recall that once upon a time, early in this century, classical physics was afflicted with ultraviolet divergences (the black body catastrophe, the electron spiraling into the nucleus, certain paradoxes in the theory of chemical reactions, etc.).  Suppose one had been told that the solution of these paradoxes is that the continuum of classical phase space, with its coordinates $p$ and $q$, does not exist, and must be replaced by a system with far fewer degrees of freedom - a system which on average has one degree of freedom per $2 \pi \hbar$ of area.  \emph{Z. Physik} might in that case have been inundated with papers proposing a phase space lattice - square lattices with lattice spacing $\sqrt{2 \pi \hbar}$, triangular lattices with lattice spacing $(8 \pi \hbar)^{1/2}/3^{1/4}$, and so on.  On the other hand, some might have argued that the fundamental phenomena would be independent of the details of the cutoff.  The correct state of affairs is of course that the actual cutoff in classical physics is a subtle one, involving Heisenberg's formula $[p,x] = -i\hbar$; and it is definitely necessary to discover this cutoff, since many of the characteristic ideas of quantum mechanics arise in answering questions that are difficult to ask until the correct cutoff is discovered.  A new version of Heisenberg's principle - some non-commutativity where it does not usually arise - may be the key to the thinning of the degrees of freedom that is needed to describe string theory correctly."}
\end{quotation}
\doublespacing
It is in this spirit that we attempt to model the holographic `thinning of degrees of freedom' via a modified commutator.  In this modification, the size of the quanta will now depend on the energy scale.  The idea of modified commutators is not new \cite{Kempf:1994qp} \cite{Scardigli:2003kr}, and has recently been employed by Kempf \cite{Kempf:2000ac} to describe string-like minimal distances in quantum field theory.  We use a similar ansatz in that we replace the spatial momentum basis $p^i$ with a more fundamental $\rho^i$-basis on which the momenta depend.  An ansatz consistent with Galilean invariance is $p^i = f(\rho) \rho^i$, where $f(\rho)$ is some function of the radial parameter $\rho  = \sqrt {\rho^i \rho^i}$.  In this basis the position operator becomes $x^i \rightarrow i \hbar \frac{\partial}{\partial \rho^i}$.  Then the commutator is
\[ [ x^i, p^j ] = i \hbar \frac{\partial p^j}{\partial \rho^i} = i \hbar \left(  f (\rho) \delta^{ij} + 2 \frac{\partial f (\rho)}{\partial \rho^2} \rho^i \rho^j \right) . \]
When $f(\rho) \equiv 1$, the usual commutator is recovered.  We stress that we have no reason for believing the commutator truly obeys this relation at a fundamental level; rather, we expect this may be some effective description which can be easily implemented to exhibit gravity-like behavior and in particular holography in \emph{any} physical system, without the need for elaborate constructions.
\section{A Simple Example}
To see how this might come about, we will consider the simplest possible example based on dimensional analysis of black holes.  In usual quantum mechanics, $p \equiv | {\bf p}| = \rho$.  Since $\rho$ is now the Fourier conjugate of position, we can interpret this to mean $p \sim \hbar/L$, which is indeed the case: larger momenta probes smaller distance.  But trying to probe ultrashort distances with $p \gg M_{pl} = \sqrt{\hbar/G}$ will be unsuccessful, since this will instead produce a black hole of size $L \sim G p$.  In our formalism this would be represented as $L \sim \hbar/ \rho \sim G p$.  These two behaviors are shown in Figure 1.  The simplest way to combine these behaviors is to use the relation 
\begin{figure}
\begin{center}
\includegraphics{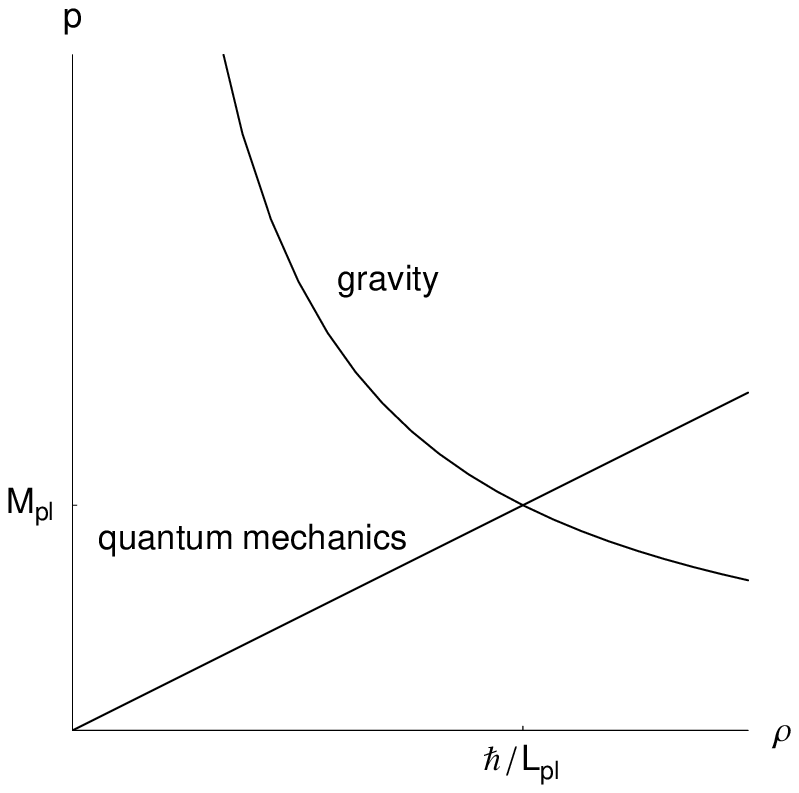}
\caption{In quantum mechanics, $p \sim \rho \sim \hbar / L$ implying large momentum probes small distance.  In black holes found in gravity, $p \sim \hbar/G\rho \sim L/G$, implying large momentum forms a large black hole.  There must be some sort of crossover behavior near the planck length $L_{pl}=\sqrt{G \hbar}$ and energy $M_{pl}=\sqrt{\hbar/G}$.} \ \\
\includegraphics{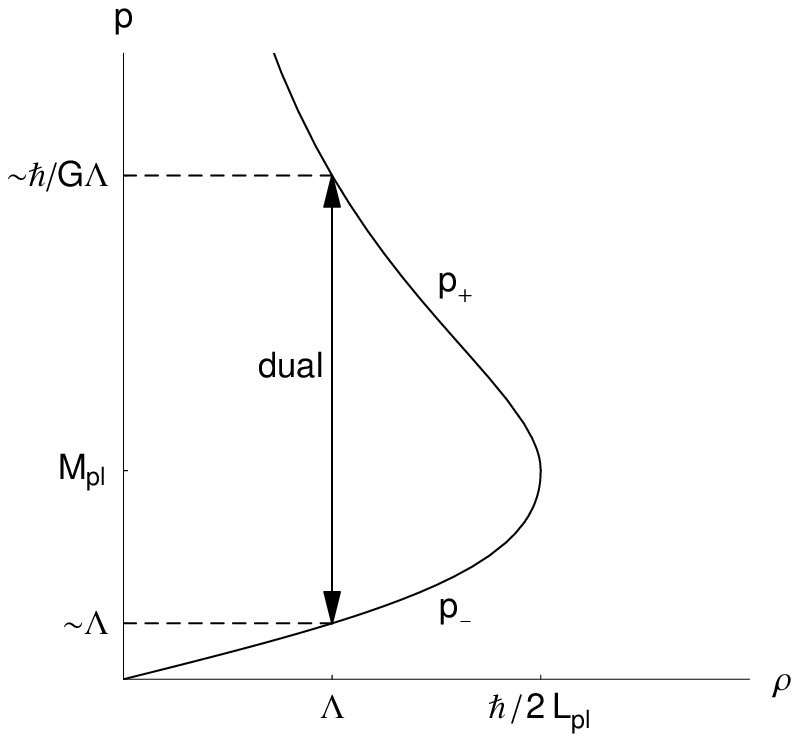}
\caption{Using the two solutions $p_\pm(\rho)$, there exist dual descriptions of the same underlying physics.  In particular, a cutoff $\rho \leq \Lambda$ implies a UV cutoff in quantum mechanics $p_- \leq p_- (\Lambda) \sim \Lambda$ as well as an IR cutoff in gravity $p_+ \geq p_+ (\Lambda) \sim \hbar / G \Lambda $. } \ \\
\end{center}
\end{figure}
\[ \rho = \frac{p}{1 + G p^2/\hbar}. \]
This can be inverted to give $p$ as a function of $\rho$,
\[  p_{\pm} = \frac{ \hbar \left( 1 \pm \sqrt{1 - 4G \rho^2/ \hbar} \right)}{2G \rho}. \]
These solutions are shown in Figure 2.  At large distance (small $\rho$) the negative solution contains the usual quantum behavior $p_- \sim \rho$, and the positive solution contains the gravitational behavior $p_+ \sim \hbar/G \rho$.  To determine the precise commutator, we solve for $f(\rho)$:
\[ f(\rho)_{\pm} = p_{\pm} /\rho = \frac{ \hbar \left( 1 \pm \sqrt{1 - 4G \rho^2 / \hbar} \right)}{2G \rho^2} . \]

Note that the solution requires $\rho \leq \sqrt{\hbar / 4G}$, implying a minimal distance $2L_{pl} =2 \sqrt{G \hbar}$ at a momentum $p = M_{pl}$.  The solution studied by Kempf also contains a short-distance cutoff, but ours is significantly different in that our minimal length can be reached at finite energy, and there is then a smooth transition to the new branch $p_+$.
\section{Duality and Dimensional Reduction}
Since there are now two values of $f$ and $p$ for each $\rho$ there is an obvious double description:  one as the low-energy ``quantum mechanical" description, the other the high-energy ``gravity" description.  The mapping between the two is very simple:  just choose the other $f_{\pm}$ and $p_\pm$ with the same $\rho$.  Note we must choose precisely one of these two descriptions;  trying to use both descriptions simultaneously is redundant and contradictory, since our wavefunction can only depend on $\rho^i$, not $p^i$.  In our example solution this duality is particularly simple to utilize, since it merely amounts to alternating between the two roots of a quadratic equation.  This QM-gravity duality makes manifest the IR/UV correspondence emphasized by Susskind and Witten \cite{Susskind:1998dq}: a cutoff $\rho \leq \Lambda$ implies a UV cutoff in quantum mechanics $p_- \leq p_- (\Lambda) \sim \Lambda$ and an IR cutoff in gravity $p_+ \geq p_+ (\Lambda) \sim \hbar / G \Lambda$. 

Let us also make a comment regarding the most natural basis to use.  Consider how the number of states of a system changes when written in terms of $\rho$ and generalizing $\hbar^3$ to be the determinant of the $3 \times 3$ commutator matrix $[x^i,p^j]$:
\[ \sum_{\rm states} = V \int \frac{d^3 p}{(2 \pi \hbar)^3} \rightarrow  V \int \frac{ d^3 p}{(2 \pi)^3 |[x^i,p^j]|} = V \int \frac{ d^3 \rho \left| \frac{\partial p^i}{\partial \rho^j} \right|}{(2 \pi \hbar)^3 \left| \frac{\partial p^i}{\partial \rho^j} \right|} = V \int \frac{d^3 \rho}{( 2 \pi \hbar)^3}. \]
The fact that the phase space is always `flat' when written in terms of $\rho^i$, regardless of the specific form of $f(\rho)$, suggests that it is $\rho^i$ and not $p^i$ that are the true degrees of freedom in the system.  What happens when we insist on using the momentum basis?  If we are on the quantum branch, the modifications to $\hbar$ would show up as additional `correction' terms in the effective action.  When we are in the gravity branch, the long-distance behavior is described by $p \sim \hbar /G \rho$, $\partial p^i / \partial \rho^j \sim -\hbar /G\rho^2$, $| \partial p^i / \partial \rho^j | \sim ( \hbar / G \rho^2)^3 \sim (G p^2 / \hbar)^3$.  \emph{Thus a system of gravity has as many states as a QM system in 6 dimensions less:}
\[ \sum_{\rm states \ in \ 3d \ gravity} = V \int \frac{d^3 p}{(2 \pi)^3 |[x^i,p^j]|} = V \int \frac{p^2 \ dp}{(2 \pi Gp^2)^{3}} \]
\[ = VM^6_{pl}  \int \frac{p^{-4} \ dp}{(2 \pi \hbar)^3} = VM^6_{pl}  \int \frac{d^{-3}p}{(2 \pi \hbar)^3} \sim  \sum_{\rm states \ in \ (-3)d \ QM}.  \]
The increase of quanta size at high energies gives an intuitive feel for why physics will prevent too many degrees of freedom in a region of space; we propose that these `fat quanta' are in fact the basic constituents of black holes.  Susskind \cite{Susskind:1994vu} has emphasized that this type of quanta growth with momenta is essential to understanding holography.  Although the amount of dimensional reduction is much larger than that found in usual holography (6 dimensions reduced instead of just 1) we remind the reader we have used only the most simplistic dimensional analysis and a commutator in non-relativistic quantum mechanics; we expect a more sophisticated treatment in gravity and quantum field theory to produce the correct amount of dimensional reduction.
\section{Conclusion}
In this essay we have presented a novel method for studying gravity and holography.  This method has the advantage that it can be implemented in any quantum system simply by changing the position-momentum commutator.  Theoretical and perhaps observational \cite{Hogan:2002xs} \cite{Hogan:2003mq} consequences could then be extracted.
\section{Acknowledgments}
This work was supported by the DOE and the NASA grant NAG 5-10842 at Fermilab.  I would like to thank C. Hogan for useful discussions, and the University of Washington-Seattle Physics and Astronomy Groups for hospitality while some of this work was being completed.

\end{document}